\documentclass[pre,floatfix,twocolumn]{revtex4}
\usepackage{amsmath}
\usepackage{amsfonts}
\usepackage{mathrsfs}
\usepackage{epsfig}
\usepackage{graphicx}
\usepackage{dcolumn}
\usepackage{amssymb}
\usepackage{hyperref}
\usepackage[hyphenbreaks]{breakurl}

\begin{document}

\title{ 
Uncovering the hierarchical structure of the international FOREX market
by using similarity metric between the fluctuation distributions of currencies}

\author{Abhijit Chakraborty, Soumya Easwaran
and Sitabhra Sinha}

\affiliation{
\begin {tabular}{c}
The Institute of Mathematical Sciences, CIT Campus, Taramani, Chennai
600113, India
\end{tabular}
}
\begin{abstract}
The decentralized international market of currency trading is
a prototypical complex system having a highly heterogeneous
composition. To understand the hierarchical structure relating the
price movement of different currencies in the market, we have focused on quantifying
the degree of similarity between the distributions of exchange rate
fluctuations. For this purpose we use
a metric constructed using the
Jensen-Shannon divergence between the normalized logarithmic return
distributions of the different currencies. This provides a novel method 
for revealing associations between currencies in terms of the statistical
nature of their rate fluctuations,
which is distinct from the conventional correlation-based methods.
The resulting clusters are consistent with the
nature of the underlying economies but also
show striking divergences during periods of major international crises. 
\end{abstract}


\maketitle

\section {Introduction}
One of the biggest challenges facing physicists trying to understand
natural and man-made complex systems is to uncover principles underlying 
their collective dynamics. Socio-economic phenomena, 
in particular, provide examples of systems that exhibit a high degree of
heterogeneity in the dynamical behavior exhibited by their components~\cite{Sinha2011}.
Unlike the dynamical systems traditionally investigated by physicists,
such as arrays of coupled oscillators, coupling between the
different parts of such complex systems do not necessarily result in 
the components displaying similar behavior (analogous to synchronization
in oscillator arrays~\cite{Pikovsky2001}). Instead, autonomous agents in economic systems may
employ various strategies which could manifest in different
segments of the system exhibiting qualitatively 
distinct behavior~\cite{Menon2019}. This is observed,
for instance, in a financial market whose
collective dynamics can be described in terms of the price movements of the
numerous assets that are traded in it. These movements are not independent
of each other as the components of this complex system ``interact'' through the actions of agents (ranging
from individual investors to large financial institutions) buying and selling these assets.
The statistical properties of the asset price fluctuations can
differ remarkably from one asset to another, which can be linked to intrinsic, as well as, extrinsic
factors. To understand the mechanisms through which such properties emerge, we first need to describe accurately
how the fluctuations of the different components are related to each other.

In this paper, we consider the largest financial market in the world,
the international currency or foreign exchange (FOREX) market~\cite{BIS2013}.
To provide a comparison, the
average daily turnover in the FOREX market was $5.3 \times 10^{12}$ USD in April 2013 
while the average daily turnover for
New York Stock Exchange, the biggest stock market in the world, in 2013 was 
$1.69 \times 10^{11}$ USD.
It is a decentralized global market for trading
currencies which plays a key role in the
international monetary and financial system. Furthermore, it is among the
freest and most competitive markets in the world. The foreign
exchange market is particularly important to complex systems researchers 
because of the large volumes of data it is able to provide for statistical analysis. 

Currency exchange rates fluctuate because of a number of reasons, including
the balance of trade, interest rate, monetary policies, etc.
As such it is very difficult to discern the individual role that
these different factors play in influencing the price fluctuations of
a currency. Instead, we propose to uncover the hierarchical structure
of the network relating the different currencies in terms of their
fluctuation behavior.
We use a metric for measuring the distance between
pairs of exchange rate fluctuation distributions to cluster the currencies 
(and by extension, the economies to which they belong) into similarity groups. This provides a novel
approach to grouping components in complex systems according to
the statistical characteristics of their dynamical behavior, distinct
from the widely employed method of using cross-correlations 
between the time-series~\cite{Laloux1999,Plerou1999a,McDonald2005,Pan2007,Kovur2014} which have
limitations~\cite{Reshef2011,Kinney2014}. We have also performed a
temporally resolved analysis of the nature of the distributions at
different periods that shows strong disruption in the otherwise regular
pattern of systematic variations during the global financial crisis of 2008,
indicating its deep-rooted nature affecting the real
economy~\cite{Claessens2012,Kenourgios2015}.
\begingroup
\squeezetable
\begin{table*}
   \caption{The currencies of developed (1-14), emerging (15-44) and
   frontier (45-75) economies considered in the study. The columns
   indicate the currency code along with the nature of the exchange rate regime (as obtained
   from Oanda and XE sites), the character of the economy (as
   categorized by MSCI) and the geographical region of the
corresponding countries.}
   \begin{ruledtabular}
   \begin{tabular}{l l l l l l}
Sl. no.  &  Currency                &  Code   & Exchange Rate Regime    &  Market Type       &  Region      \\
 &  &  &  (Oanda, XE) & (MSCI)  &   \\ \hline \hline
1 &  Canadian Dollar                &  CAD    & Floating              &  Developed         & Americas     \\
2 &  Danish Krone                   &  DKK    & Pegged within horizontal band         &  Developed         & Europe       \\
3 &  Euro                           &  EUR    & Floating              &  Developed         & Europe    \\
4 &  Great Britain Pound            &  GBP    & Floating              &  Developed         & Europe     \\
5 &  Iceland Krona                  &  ISK    & Floating              &  Developed         & Europe   \\
6 &  Norwegian Kroner               &  NOK    & Floating              &  Developed         & Europe    \\
7 &  Swedish Krona                  &  SEK    & Floating              &  Developed         & Europe       \\
8 &  Swiss Franc                    &  CHF    & Floating              &  Developed         & Europe       \\
9 &  Israeli New Shekel             &  ILS    & Floating              &  Developed         & Middle East  \\
10 &  Australian Dollar             &  AUD    & Floating              &  Developed         & Asia-Pacific    \\
11 &  Hong Kong Dollar              &  HKD    & Fixed peg         &  Developed         & Asia-Pacific \\
12 &  Japanese Yen                  &  JPY    & Floating              &  Developed         & Asia-Pacific  \\
13 &  New Zealand Dollar            &  NZD    & Floating              &  Developed         & Asia-Pacific     \\
14 &  Singapore Dollar              &  SGD    & Floating              &  Developed         & Asia-Pacific        \\
\hline
15 &  Bolivian Boliviano            &  BOB    & Crawling peg              &  Emerging          & Americas           \\
16 &  Brazilian Real                &  BRL    & Floating              &  Emerging          & Americas           \\
17 &  Chilean Peso                  &  CLP    & Floating              &  Emerging          & Americas           \\
18 &  Colombian Peso                &  COP    & Floating              &  Emerging          & Americas         \\
19 &  Dominican Republic Peso       &  DOP    & Floating              &  Emerging          & Americas        \\
20 &  Mexican Peso                  &  MXN    & Floating              &  Emerging          & Americas         \\
21 &  Peruvian Nuevo Sol            &  PEN    & Floating              &  Emerging          & Americas        \\
22 &  Venezuelan Bolivar            &  VEB    & Fixed peg         &  Emerging          & Americas           \\
23 &  Albanian Lek                  &  ALL    & Floating              &  Emerging          & Europe               \\
24 &  Czech Koruna                  &  CZK    & Floating              &  Emerging          & Europe          \\
25 &  Hungarian Forint              &  HUF    & Pegged within horizontal band              &  Emerging          & Europe      \\
26 &  Polish Zloty                  &  PLN    & Floating              &  Emerging          & Europe      \\
27 &  Russian Rouble                &  RUB    & Floating              &  Emerging          & Europe              \\
28 &  Turkish Lira                  &  TRY    & Floating              &  Emerging          & Europe           \\
29 &  Algerian Dinar                &  DZD    & Floating              &  Emerging          & Africa          \\
30 &  Cape Verde Escudo             &  CVE    & Fixed peg         &  Emerging          & Africa          \\
31 &  Egyptian Pound                &  EGP    & Floating              &  Emerging          & Africa       \\
32 &  Ethiopian Birr                &  ETB    & Floating              &  Emerging          & Africa            \\
33 &  Mauritius Rupee               &  MUR    & Floating              &  Emerging          & Africa           \\
34 &  Moroccan Dirham               &  MAD    & Fixed peg              &  Emerging          & Africa           \\
35 &  South African Rand            &  ZAR    & Floating              &  Emerging          & Africa            \\
36 &  Tanzanian Shilling            &  TZS    & Floating              &  Emerging          & Africa            \\
37 &  Chinese Yuan Renminbi         &  CNY    & Fixed peg             &  Emerging          & Asia               \\
38 &  Indian Rupee                  &  INR    & Floating              &  Emerging          & Asia                 \\
39 &  Indonesian Rupiah             &  IDR    & Floating              &  Emerging          & Asia              \\
40 &  Papua New Guinea Kina         & PGK  & Floating                 &  Emerging          & Asia             \\
41 &  Philippine Peso               &  PHP    & Floating              &  Emerging          & Asia              \\
42 &  South Korean Won              &  KRW    & Floating              &  Emerging          & Asia            \\
43 &  Taiwan Dollar                 &  TWD    & Floating              &  Emerging          & Asia              \\
44 &  Thai Baht                     &  THB    & Floating              &  Emerging          & Asia               \\
\hline
45 &  Guatemalan Quetzal            &  GTQ    & Floating              &  Frontier          & Americas         \\
46 &  Honduran Lempira              &  HNL    & Crawling peg              &  Frontier          & Americas           \\
47 &  Jamaican Dollar               &  JMD    & Floating              &  Frontier          & Americas         \\
48 &  Paraguay Guarani              &  PYG    & Floating              &  Frontier          & Americas         \\
49 &  Trinidad Tobago Dollar        &  TTD    & Floating              &  Frontier          & Americas            \\
50 &  Croatian Kuna                 &  HRK    & Floating              &  Frontier          & Europe              \\
51 &  Kazakhstan Tenge              &  KZT    & Floating              &  Frontier          & Europe               \\
52 &  Latvian Lats                  &  LVL    & Fixed peg         &  Frontier          & Europe            \\
53 &  Botswana Pula                 &  BWP    & Crawling peg              &  Frontier          & Africa        \\
54 &  Comoros Franc                 &  KMF    & Fixed peg         &  Frontier          & Africa          \\
55 &  Gambian Dalasi                &  GMD    & Floating              &  Frontier          & Africa         \\
56 &  Ghanaian Cedi                 &  GHC    & Floating              &  Frontier          & Africa               \\
57 &  Guinea Franc                  &  GNF    & Fixed peg              &  Frontier          & Africa               \\
58 &  Kenyan Shilling               &  KES    & Floating              &  Frontier          & Africa               \\
59 &  Malawi Kwacha                 &  MWK    & Floating              &  Frontier          & Africa                \\
60 &  Mauritanian Ouguiya           &  MRO    & Floating              &  Frontier          & Africa              \\
61 &  Mozambique Metical            &  MZM    & Floating              &  Frontier          & Africa               \\
62 &  Nigerian Naira                &  NGN    & Floating              &  Frontier          & Africa                 \\
63 &  Sao Tome and Principe Dobra   &  STD    & Fixed peg         &  Frontier          & Africa                 \\
64 &  Zambian Kwacha                &  ZMK    & Floating              &  Frontier          & Africa               \\
65 &  Jordanian Dinar               &  JOD    & Fixed peg         &  Frontier          & Middle East         \\
66 &  Kuwaiti Dinar                 &  KWD    & Fixed peg              &  Frontier          & Middle East       \\
67 &  Syrian Pound                  &  SYP    & Fixed peg             &  Frontier          & Middle East          \\
68 &  Brunei Dollar                 &  BND    & Fixed peg  &  Frontier          & Asia        \\
69 &  Bangladeshi Taka              &  BDT    & Floating              &  Frontier          & Asia            \\
70 &  Cambodian Riel                &  KHR    & Floating              &  Frontier          & Asia             \\
71 &  Fiji Dollar                   &  FJD    & Fixed peg              &  Frontier          & Asia              \\
72 &  Lao Kip                       &  LAK    & Floating              &  Frontier          & Asia               \\
73 &  Pakistan Rupee                &  PKR    & Floating              &  Frontier          & Asia                 \\
74 &  Samoan Tala                   &  WST    & Fixed peg              &  Frontier          & Asia                \\
75 &  Sri Lankan Rupee              &  LKR    & Floating   &  Frontier          & Asia         \\ 
\end{tabular}
\end{ruledtabular}
\label{table1}
\end{table*}
\endgroup

\section {Materials and Methods}
\label{DataMethods}
		
In this study we have analyzed a data-set comprising the daily
exchange rates of $N=75$
currencies (see Table~\ref{table1}) with respect to the US Dollar (USD), 
for the period October 23, 1995 to February 10, 2016,
corresponding to $\tau=7416$ days.
Specifically, we use the midpoint between the bid and ask rates for 1 USD against a given currency. 
The source of the data is an online archive of historical interbank market 
rates made available by the Oanda Corporation~\cite{Oanda} which provides
internet-based currency trading and information services.
The interbank rate (also referred to as spot rate) is used for large
volume transactions carried out by banks and financial institutions,
and is the exchange rate that is typically reported in the media.
From the online archive we have obtained the rate for each day in the
period under consideration, averaged over all quotations
collected within the previous 24 hour period (terminating at midnight UTC) 
from various frequently updated sources around the world, 
including online currency trading platforms, market data
vendors, and contributing financial institutions.
The USD is chosen as the numeraire as it is
the preferred currency for most international transactions and
continues to remains the reserve currency of choice for many
economies~\cite{Papaioannou2006,IMFAnnRep}.

To see how the use of a particular numeraire, viz., the USD, may have introduced
bias in the results reported here, we have also carried out our analysis by expressing
the exchange rates of currencies in terms of Special Drawing Rights (SDR), a supplementary
international reserve asset created by the International Monetary Fund (IMF), whose value
is determined by a basket of currencies~\cite{SDR}. The composition of the basket 
is altered regularly to ensure that it reflects the relative importance of different currencies
in international trade and finance. Daily quotes for SDR
in terms of USD is available from the IMF~\cite{IMFdata}
for all days (except those on which IMF is closed for business, which include weekends).
We have used this to create a daily exchange rate time-series with SDR as the
numeraire, spanning the $5210$ days
for which data is available in the period under study.

The $75$ currencies considered in our study are chosen based on 
the exchange rate regime followed by them (see Table 1). The information
about the regime type of each currency has been obtained from the same
online source from which we have collected the daily rates~\cite{Oanda},
supplemented with information from the website of another online 
foreign exchange tools and services company~\cite{xe}. 
We have explicitly avoided any currency whose exchange rate with respect 
to USD does not vary over time. Most of the currencies chosen for our analysis
are either freely floating
under the influence of market forces or managed to an extent with no
pre-determined path. The remainder, although pegged
to USD or some other important currency (such as EUR), do exhibit
variation within a band which may be either fixed or can change over
time. 
We have also classified the economy of the countries to which these
currencies belong by using the Morgan Stanley Capital 
International (MSCI) market classification framework~\cite{MSCI}.
Using multiple criteria including the sustainability of economic 
development, the number of companies meeting certain size and 
liquidity criteria, ease of capital flow, as well as, efficiency and 
stability of the institutional framework, MSCI classifies the
economies into three categories,
viz., developed, emerging and frontier markets (see Table 1).
An important point to note is that for the period prior to  January 
1, 1999, when the EUR was first introduced, we have used in its stead 
the exchange rate for the ECU (European Currency Unit).

\section{Results}

Instead of analyzing the raw price information, the fluctuations in the exchange rates of
75 currencies (see Table 1) with respect to the US Dollar
over the period 1995-2016 is measured. To make the result independent of the unit of measurement,
the variation in the exchange
rate $P_i(t)$ of the $i$-th currency ($i = 1, \ldots, N$) at time $t$ is quantified 
by its logarithmic return defined over a time-interval $\Delta t$ as 
$R_i(t,\Delta t) = ln \, P_i(t+\Delta t)-ln\, P_i(t) $. 
Since our data comprises the daily exchange rates, we consider $\Delta t = 1$ day. 
The standard deviation $\sigma$ of the returns measures the intensity of fluctuations in the exchange
rates (volatility) which varies for different currencies. Hence, to compare between
the return distributions of different currencies, we normalize the
returns of each currency $i$ by subtracting the mean value $\langle
R_i \rangle = \Sigma_{t=1}^T R_i (t)/T$ and
dividing by the standard deviation $\sigma_i (t) =\sqrt{\frac{1}{T-2}
\Sigma_{t^\prime \neq t} [R_i (t^\prime) - \langle R_i \rangle]^2}$
(removing the self contribution from the measure of 
volatility), obtaining
the normalized return, $ r_i(t)=(R_i(t)-\langle R_i
\rangle)/\sigma_i(t)$.

\subsection{Hierarchical clustering based on similarity of
fluctuations distribution}
\begin{figure}[tbp]
\begin{center}
\includegraphics[width=0.99\linewidth,clip]{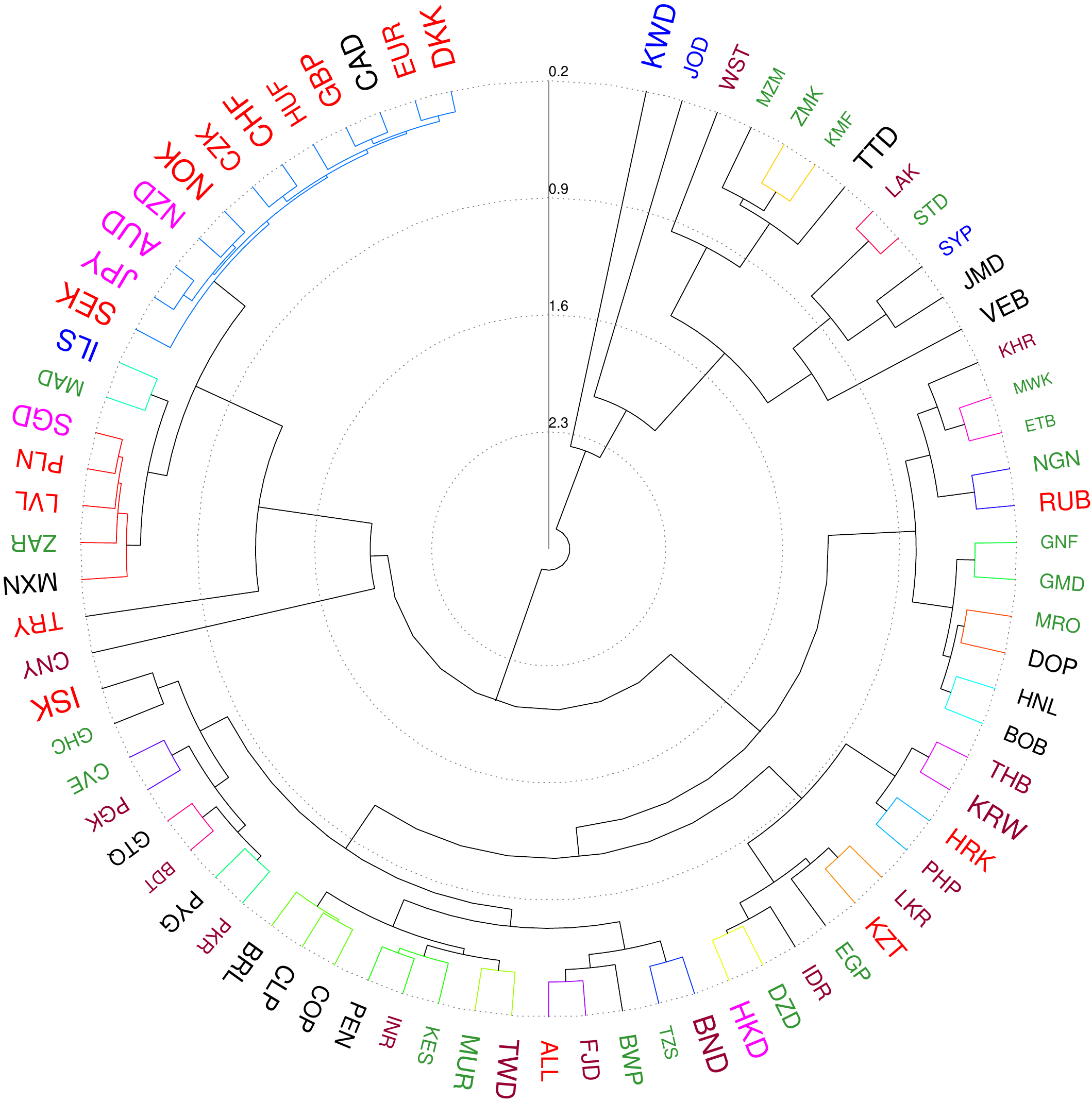}
\end{center}
\caption{
{\bf Polar dendrogram representation obtained by hierarchical clustering
of different
currencies in terms of
the statistical distance between their fluctuation distributions.}
A similarity distance $D$ obtained from the Jensen-Shannon divergence 
between the corresponding
normalized logarithmic return distributions of a pair of currencies
has been used as the clustering metric.
The currencies have been clustered using complete linkage algorithm
and the height of a branch
measures the linkage function $d$, i.e., the distance between two
clusters.
Using a threshold of $d_{th} = 0.47$, the largest number of distinct
clusters
(viz., 22 clusters represented by the different colored branches of
the dendrogram,
black branches indicating isolated nodes) can be 
identified. The largest cluster comprises only currencies of
developed economies with the exception of CZK and HUF (belonging to emerging
economies).
Currencies are distinguished according to the average annual GDP per
capita $\langle g \rangle$
of the corresponding economy (represented by font size, which scales
logarithmically with $\langle g \rangle$) and the geographical region to which they
belong (represented by font color, viz., black: Americas, red: Europe,
blue: Middle East, magenta: Asia-Pacific, green: Africa and brown:
Asia).
}
\label{fig5}
\end{figure}
\begin{figure*}[h]
\begin{center}
\includegraphics[width=0.99\linewidth,clip]{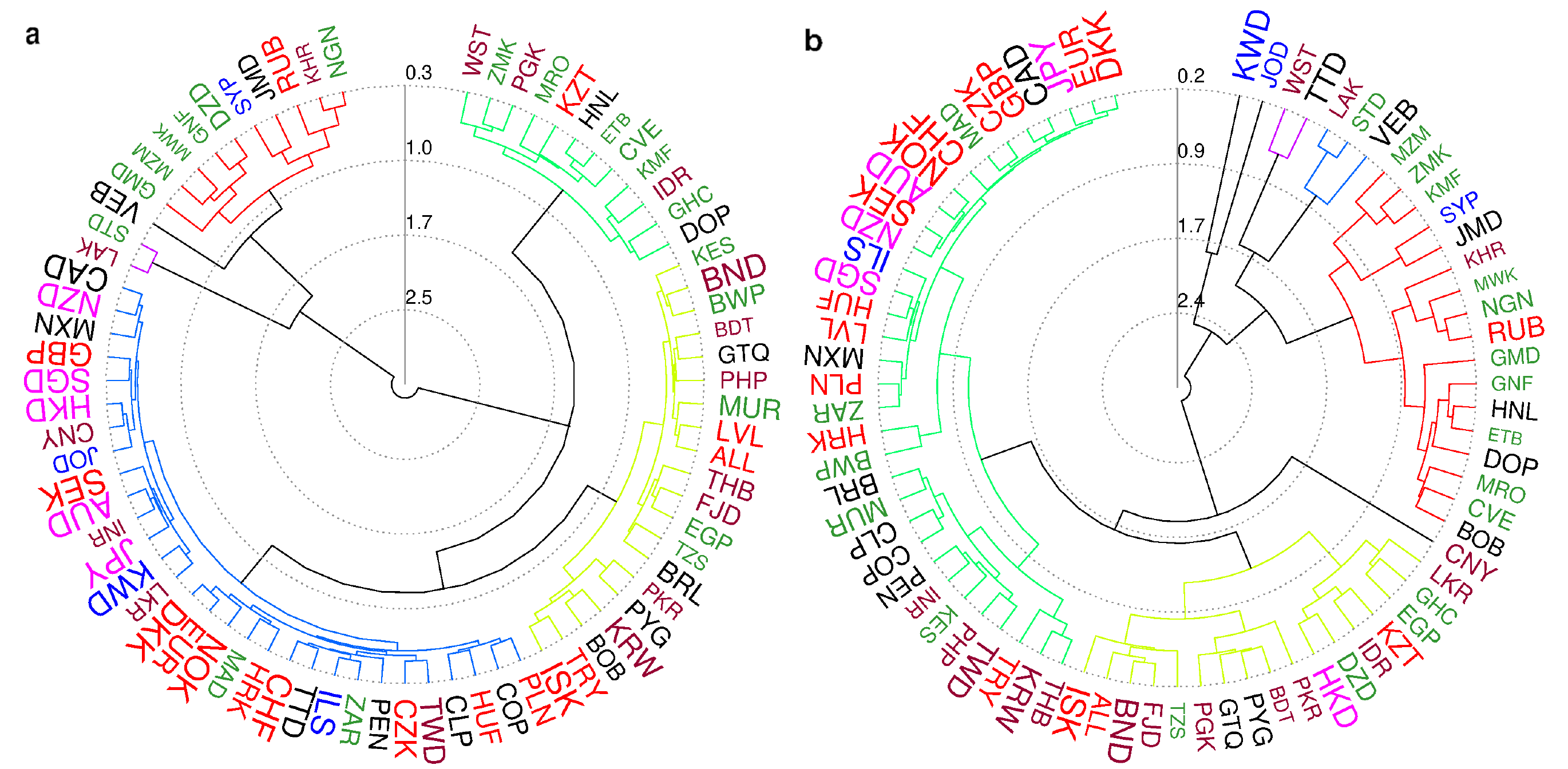}
\end{center}
\caption{{\bf Comparison between hierarchical clustering of currencies according to the statistical
distance between corresponding fluctuation distributions when the
exchange rates are expressed in terms of different numeraires, viz., (a) SDR and (b) USD.}
As in Fig.~\ref{fig5}, clustering is performed using complete linkage 
algorithm.
In (a), using a threshold of $d_{th} = 0.78$, $5$ distinct
clusters, represented by the different colored branches of
the dendrogram (with
black branches indicating isolated nodes), can be 
identified. Note that the SDR data used for constructing the dendrogram is available only for $5210$ days during the period
October 23, 1995 to February 10, 2016. Using this same set of dates, we have constructed in (b) a dendrogram from the data
in which currency exchange rates are expressed in terms of USD, for comparison with (a). For a
threshold of $d_{th} = 1.30$, we observe $5$ distinct clusters in this case also. The cophenetic distance correlation coefficient between 
the two dendrograms is $0.52$, suggesting a high degree of similarity between them.
}
\label{figSDR}
\end{figure*}
\begin{figure}[h]
\begin{center}
\includegraphics[width=0.99\linewidth,clip]{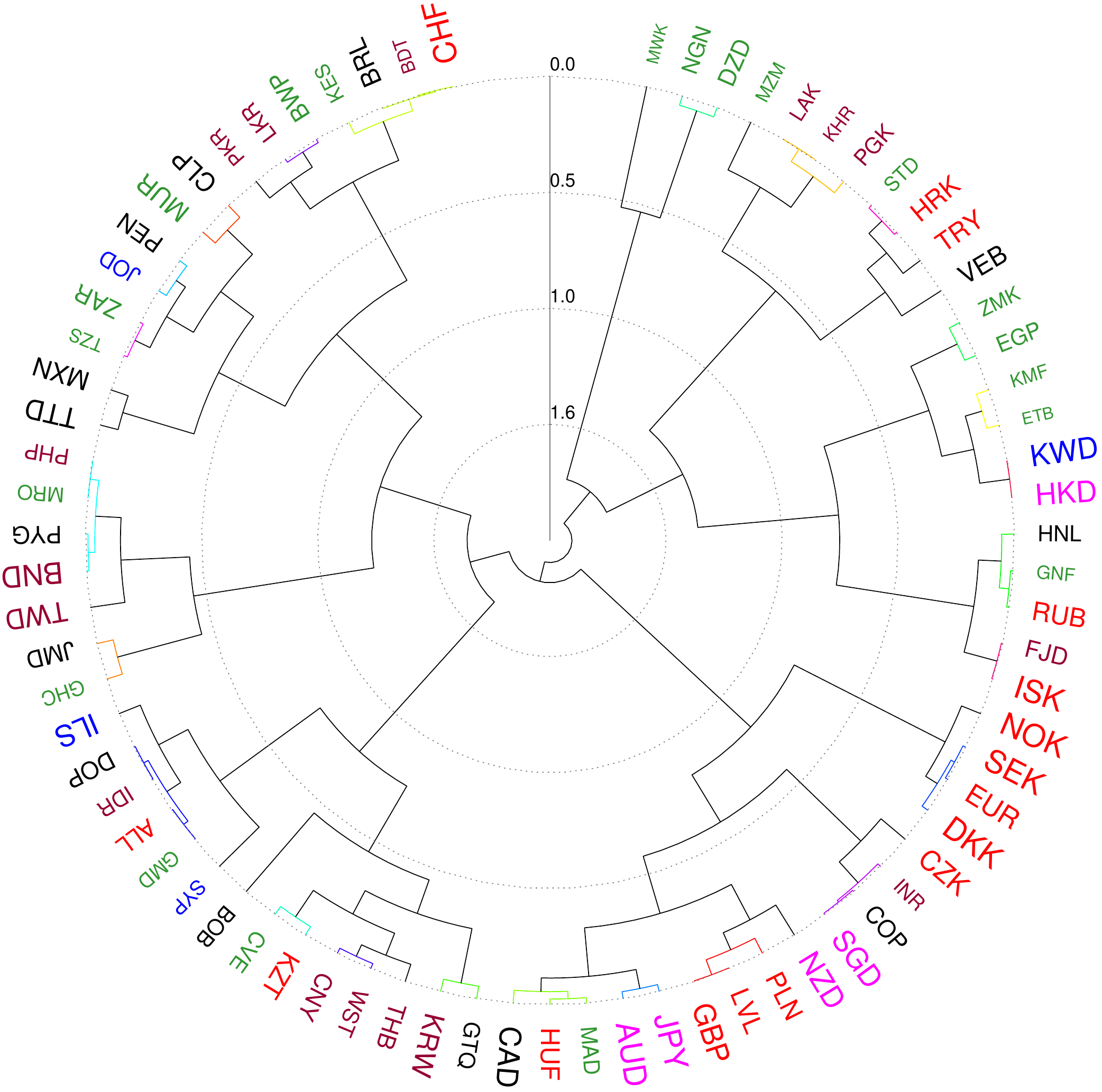}
\end{center}
\caption{
{\bf Polar dendrogram representation obtained by hierarchical clustering
of different currencies on the basis of how
similar they are in exhibiting extreme values in their exchange rates,
i.e., in terms of the heavy-tailed nature of the fluctuation distributions.}
The relative difference between the kurtosis of corresponding
normalized logarithmic return distributions of a pair of currencies
has been used as the clustering metric.
As in Fig.~\ref{fig5}, clustering is performed using complete linkage 
algorithm.
Using a threshold of $d_{th} = 0.08$, the largest number of distinct
clusters (viz., 24, represented by the different colored branches of
the dendrogram, with
black branches indicating isolated nodes) can be 
identified. Unlike the clusters obtained using 
Jensen-Shannon divergence shown in Fig.~\ref{fig5}, all clusters here are 
relatively small. 
The relative dissimilarity between the clusters seen here and that in Fig.~\ref{fig5} is quantitatively
indicated by the value of the cophenetic distance correlation coefficient between the two dendrograms being $0.2$.   
However, most of the currencies belonging to developed economies 
do seem to occur close to each other, with the exceptions of
CHF, HKD and ISK (resulting from rare instances of extremely large
deviations in their exchange rates).
As in Fig.~\ref{fig5}, currencies
are distinguished according to the average annual GDP per
capita $\langle g \rangle$
of the corresponding economy (represented by font size, which scales
logarithmically with $\langle g \rangle$) and the geographical region to which they
belong (represented by font color, viz., black: Americas, red: Europe,
blue: Middle East, magenta: Asia-Pacific, green: Africa and brown:
Asia).
}
\label{fig6}
\end{figure}

We have investigated the inter-relation between the different
currencies by considering how similar they are in terms of the nature of
their fluctuations. For this we have measured the difference between
the normalized logarithmic return distributions of each pair of
currencies using a probability distance metric, viz., the
similarity distance $D(P_i,P_j)$ between a pair of return distributions
$P_i(r)$ and $P_j(r)$~\cite{Endres2003}. It is defined as 
the square root of the
Jensen-Shannon (JS) divergence~\cite{Lin1991},
which in turn can be defined in
terms of the Kullback-Leibler (KL) divergence for a pair of
probability distributions $P_i(x)$ and $P_j(x)$ of a discrete random 
variable $x$:
$$KL(P_i,P_j)=\sum_{x \in X}P_i(x)log\frac{P_i(x)}{P_j(x)}.$$
The limitations of the KL divergence, viz., that it is asymmetric and is, moreover,
undefined when either $P_i$ or $P_j$ is zero for any value of $x \in
X$, is overcome by the JS divergence which is defined as:
$$JS(P_i,P_j)=\frac{1}{2}KL(P_i,P)+\frac{1}{2}KL(P_j,P),$$ 
where, $P=(P_i+P_j)/2$.
As returns are continuous variables, in order to calculate the
divergences between their distributions we have discretized the
values using a
binning procedure (involving
intervals of width $\Delta r \sim 0.05$). The same binning
procedure has been also used for the analysis described later
in which the data is split into four non-overlapping segments
belonging to different time periods. Note that,
the related generalized JS measure has been used earlier 
to measure the similarity of tick frequency spectrograms for 
different currency exchange rates~\cite{Sato}. 

The matrix of similarity distances $\cal D$ between all pair of
currencies is used for clustering them in a hierarchical manner. 
Note that this approach is distinct from earlier attempts of
hierarchical classification which use synchronous cross-correlation between
exchange rate fluctuations in order to identify
clusters of related currencies (see, e.g., Refs.~\cite{Mizuno2006,Kwapien2009,Keskin2011}).
Given a set of nodes to be clustered and a matrix specifying the
distances between them, the method of hierarchical 
clustering~\cite{Johnson1967} involves (i) initially considering each
node as a cluster, (ii) merging the pair of clusters which have the
shortest distance between them, (iii) re-computing the distance
between all clusters, and repeating the steps (ii) and (iii) until all
nodes are merged into a single cluster. Clustering methods can differ
in the way the inter-cluster distance is calculated in step (iii). 
If this distance is taken as the maximum of the pairwise distances
between members of one cluster to members of the other cluster, it is
known as {\em complete-linkage} clustering. On the other hand, in the
{\em single-linkage} or nearest neighbor clustering, the
minimum of the distance between any
member of one cluster to any member of the other cluster is chosen.
Average-linkage clustering, as
the name implies, considers the mean of the pairwise distances between
members of the two clusters. Note that, the hierarchical clustering
obtained using the complete-linkage method will be same as one
obtained using a threshold
distance to define membership of a cluster, while that constructed
using the
single-linkage method is identical to the minimal
spanning tree~\cite{Gower1969}. 

We have shown the hierarchical clustering (using complete-linkage
clustering) of the different currencies
considered in this study in Fig.~\ref{fig5} using a polar dendrogram
representation. 
We note that the technique divides the currencies at the coarsest
scale into two groups, the smaller of which is almost exclusively composed of
currencies from frontier economies (the sole exception being
VEB which belongs to an emerging economy) that are characterized by large
fluctuations. Although some of these currencies (e.g., KWD, TTD and VEB) 
belong to countries having high GDP per capita, they typically 
also have a high Theil index~\cite{Theil1967} indicating that their 
foreign trade is dominated by the
export of a few key products (e.g., crude oil in the case of Kuwait,
Trinidad \& Tobago and Venezuela). Their currencies are therefore
potentially highly susceptible to fluctuations in the worldwide
demand. 
Focusing now on the larger group, we observe that it is further
divided broadly into two subgroups, one of which is dominated by
relatively stable currencies from
developed and emerging economies, with only a single frontier
economy being represented (viz., LVL) which has a
relatively higher GDP per capita than the other frontier economies.
The other subgroup is composed of currencies from
emerging as well as frontier economies (with the exception of HKD and ISK
which belong to developed economies). The occurrence of these latter two
in this subgroup can be related to the severe financial crises encountered
by these economies at different times, viz., the 2003 SARS crisis in 
the case of Hong Kong and
the 2008 banking collapse in the case of Iceland, during the period
under consideration.
The largest number of similarity clusters into which
the currencies can be grouped is obtained for
a threshold value of $d_{th}=0.47$. Most of the currencies belonging to
developed economies are seen to occur in the largest cluster consisting 
of 12 currencies,
indicating that these economies have a relatively similar high degree of
stability for their currency exchange rates. As expected, almost all of 
them have high GDP per capita. The only members
of this highly stable currency cluster which are not in the developed 
category (viz., CZK and HUF) belong to countries that are
members of the European Union, and whose economies are therefore 
highly integrated with the other members of this cluster. 
The second largest cluster comprising $5$ currencies also appear
to bring together economies that have similar GDP per
capita even though they vary in their market classification from
developed (SGD) to emerging (MXN, PLN and ZAR) and frontier (LVL).
The remaining clusters are mostly pairs (or at most triplets) of currencies.
%
%

To ensure that the hierarchical organization evident from the dendrogram is not too sensitively 
dependent on the specific numeraire we have used, in Fig.~\ref{figSDR} we have compared dendrograms
constructed using exchange rates expressed in terms of SDR and USD. As the daily quotes for SDR is available
only for a subset of the dates for which we have information about the rates expressed in USD, for a fair comparison
we have constructed the two dendrograms using the set of dates common between the two classes of data. 
As is evident, the clusters of currencies observed in the two hierarchical structures are qualitatively similar.
The resemblance between the two dendrograms can be quantified by using the cophenetic distance correlation coefficient 
(CDCC). This is computed by measuring the {\em cophenetic distance} between pairs of elements, which is defined as the height of the branch 
where two elements become members of the same cluster in the dendrogram, and
then obtaining the correlation between the corresponding distances in the two dendrograms~\cite{Sokal1962,Rohlf1981,Saracli2013}. 
The value of CDCC between the dendrograms shown in panels (a) and (b) of Fig.~\ref{figSDR} is found to be 0.52, suggesting
a high degree of similarity between them.

As the shape of a distribution can be described by its 
moments~\cite{Gavriliadis2009}, to delve deeper into whether any 
specific property of the fluctuation
distribution is responsible for the hierarchical clustering reported
above, we have investigated the relation between the lowest order
moments of the distributions for the different currencies.
As we consider normalized returns, the first and second moments (corresponding
to mean and variance) of all the distributions are identical. Furthermore,
as most of the return distributions are approximately symmetric (large
deviations being shown only by NGN, DZD, MWK, PGK, VEB, MZM, TRY, FJD and STD), the distributions
do not also markedly differ in terms of their skewness
(related to the third moment).
Thus, we focus on the kurtosis, measured in terms of the fourth moment, 
which quantifies the probability of occurrence of extreme values reflected
in the heavy-tailed nature of the distribution.
Considering higher order moments is not likely to provide any further
information as it is known that specifying a finite number of moments
of a distribution only determines its tail but not its bulk~\cite{Lindsay2000}.

\begin{figure}[tbp]
\begin{center}
\includegraphics[width=0.99\linewidth,clip]{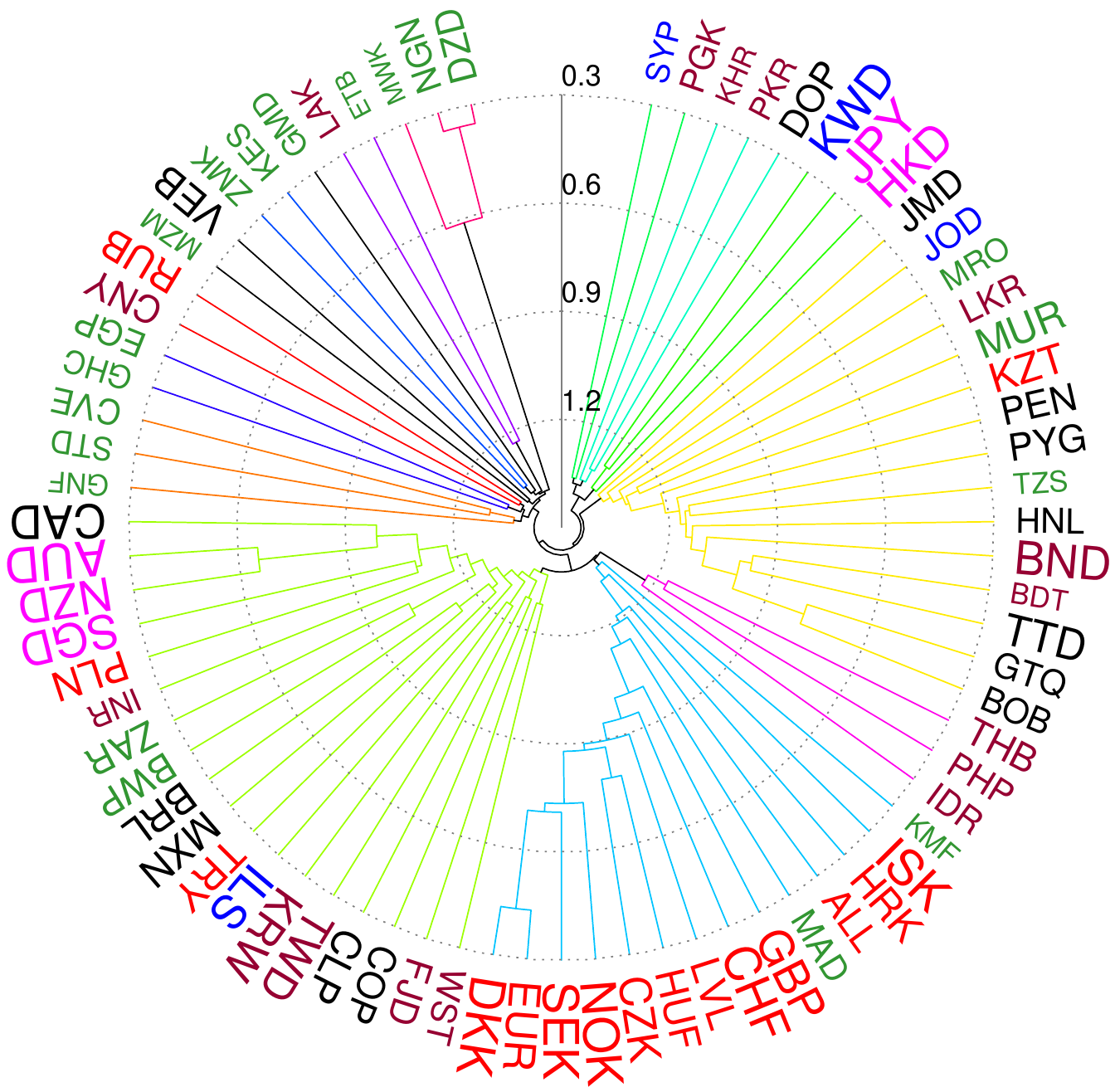}
\end{center}
\caption{
{\bf Polar dendrogram representation of the hierarchical clustering of currencies using synchronous cross-correlations between
the time-series of their exchange rate fluctuations.}  Distance between each pair of currencies $i,j$ is computed from the
Pearson correlation $C_{ij}$ between their normalized returns over the period under study as $d_{ij} = \sqrt{2 (1-C_{ij})}$, following
Ref.~\cite{Mantegna1999}. 
As in Fig.~\ref{fig5}, clustering is performed using complete linkage 
algorithm. For
a threshold of $d_{th} = 1.4$, $13$ distinct
clusters represented by the different colored branches of
the dendrogram (with
black branches indicating isolated nodes) can be 
identified. Note the cluster of three currencies (comprising DZD, NGN and MWK) at the top. Their
apparently strong association arises from a single extreme fluctuation event that occurred on November 7, 1995.  
}
\label{fig_correlation}
\end{figure}

For each pair of currencies $i, j$, we have measured the relative 
difference between the kurtosis $\alpha_4$ of their fluctuation
distributions as $$\delta(\alpha_4^i,\alpha_4^j)
=\frac{|\alpha_4^i - \alpha_4^j|}
{\langle \alpha_4^i,\alpha_4^j \rangle},$$ 
where $\langle \ldots \rangle$ represents
the mean. The matrix of these differences 
is then used to hierarchically cluster the currencies using the 
complete-linkage method. Fig.~\ref{fig6} shows the results of the
clustering using a polar dendrogram representation.
There are clearly discernible differences with the structure shown in
Fig.~\ref{fig5}, indicating that there are aspects of the distribution
other than kurtosis (and hence, the behavior at the tails) which 
are responsible for the similarity in the nature of fluctuations of 
the currencies belonging to a cluster. The differences can partly
be ascribed to the fact that the kurtosis is strongly affected by 
rare (or even single instances of) extreme events. For example, this
is possibly the explanation for the adjacency of currencies DZD, NGN
and MWK seen in Fig.~\ref{fig6}, as all three experienced a large 
deviation in their exchange rates on November 7, 1995. 
This has been explicitly verified by
reconstructing the dendrogram after discarding this specific extreme
return value and noting that the three currencies no longer occur
in close proximity. 
We also note that this single extreme event strongly affects the synchronous cross-correlations between 
the currency movements over the entire period under study, resulting in the three
currencies being identified as a prominent cluster moving independently of other currencies~\cite{Kovur2014}. 
This is apparent from the dendrogram shown in Fig.~\ref{fig_correlation} which represents relations
between the currencies in terms of the degree of synchrony in their fluctuations. It underlines one of the weaknesses of using cross-correlations
between asset price fluctuations for inferring significant dynamical relationships between the
assets, as a lone outlier may be sufficient to significantly bias the results obtained.

Another example of a rare event influencing the kurtosis is the occurrence of CHF (belonging
to a developed economy) far from the neighborhood of the other
currencies of developed economies. This can be traced to a rapid rise
in its exchange rate immediately
following the depegging of CHF from EUR in January 2015.
If the clustering is performed by considering only the time period
preceding this event, we note that CHF occurs along with the other
currencies belonging to economies having similar character.
We do note, however, that, in general, currencies belonging to similar
economies tend to be grouped in neighboring locations in the dendrogram,
which could be related to the fact that the kurtosis of the fluctuation
distribution of a currency is related to the overall prosperity,
as measured by the gross domestic product per capita, of the
corresponding economy~\cite{Chakraborty2018}.

\subsection{Temporal evolution of the system properties}

\begin{figure*}[h]
\begin{center}
\includegraphics[width=0.99\linewidth,clip]{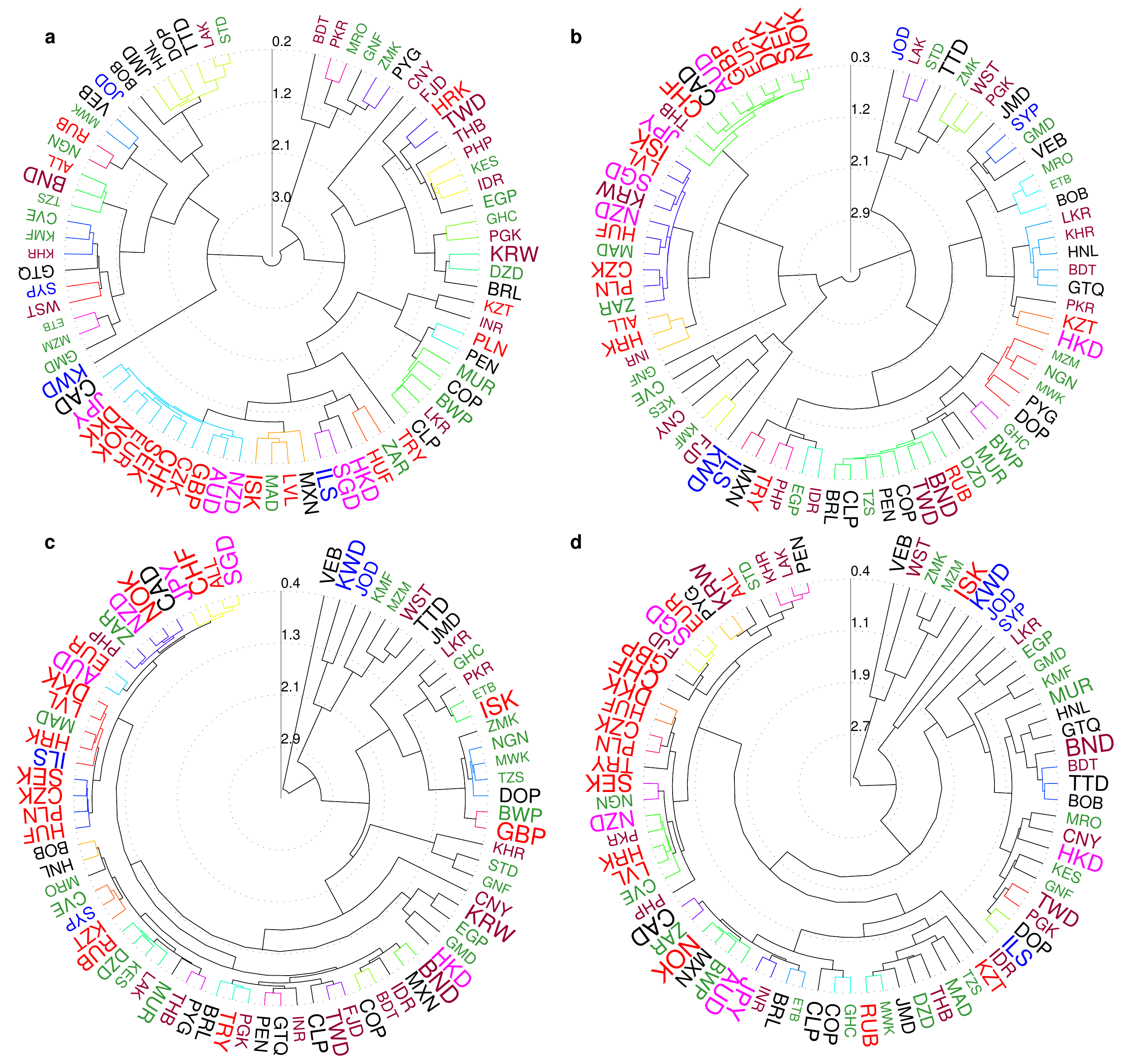}
\end{center}
\caption{
{\bf Temporal evolution of the hierarchical clustering of different
currencies according to the similarity of their exchange rate
fluctuation distributions.}
Dendrogram representations of the clusters constructed following the
same procedure as for Fig.~\ref{fig5} are
shown for four different time intervals: (a)
Period I: Oct 23, 1995-Nov 18, 2000, (b) Period II: Nov 19, 2000-
Dec 16, 2005, (c) Period III: Dec 17, 2005-Jan 13, 2011, and
(d) Period IV: Jan 14, 2011-Feb 10, 2016.
The threshold distances used for obtaining the largest number of
clusters in each period are $d_{th} = 0.80$, $0.90$, $0.70$ and $0.65$,
which yield $19$, $16$, $17$ and $16$ clusters, respectively.
Note that the statistical distance between the currencies shrink
in the latter two periods
following the major economic crisis that occurred in 2008. The cluster
of currencies belonging to developed economies seen in the earlier
periods is also seen to be disrupted to an extent.
The degree of similarity between the different dendrograms is quantified by the
cophenetic distance correlation coefficient as follows: CDCC(I,II) = $0.23$,
CDCC(II,III)=$0.12$ and CDCC(III,IV)=$0.46$. 
As in Fig.~\ref{fig5}, currencies are distinguished according to the
average annual GDP per capita $\langle g \rangle$ of the corresponding economy
(represented by font size, which scales logarithmically with $\langle
g \rangle$) and
the geographical region to which they belong (represented by font 
color, viz., black: Americas, red: Europe, blue: Middle East, magenta:
Asia-Pacific, green: Africa and brown: Asia).
}
\label{fig7}
\end{figure*}
In the analysis presented above we have considered the entire temporal duration
which our data-set spans. However, as the world economy underwent
significant changes during this period, most notably, the global
financial crisis of 2008, it is of interest to see how the
hierarchical structure of the currency market has evolved over time. 
For this purpose we
divide the data-set into four equal non-overlapping periods
comprising 1853 days,
corresponding to Period I: Oct 23, 1995 - Nov 18, 2000, Period II:
Nov 19, 2000 - Dec 16, 2005, Period III: Dec 17, 2005 - Jan 13, 2011 
and Period IV: Jan 14, 2011 - Feb 10, 2016.
Note that Period III corresponds to the crisis of the global
economy spanning 2007-2009 that also affected the FOREX market~\cite{Melvin2009}, allowing us to contrast the structure
of the currency market in the pre-crisis era (Periods I and II) with
the situation that prevailed during the crisis, as well as, post-crisis 
(Period IV).
For each of these periods, we carry out the same procedures as outlined earlier
for the entire data-set.

The hierarchical clustering of the currencies show striking
changes over time when they are constructed separately for each of the
four intervals mentioned above (Fig.~\ref{fig7}). The degree of similarity 
between the clusters seen in the four periods can be quantitatively indicated
by the cophenetic distance correlation coefficient calculated for pairs of 
successive periods.The corresponding values, viz., CDCC(I,II) = $0.23$,
CDCC(II,III)=$0.12$ and CDCC(III,IV)=$0.46$, suggest that a major
disruption took place in the hierarchical structure in the period immediately leading upto the crisis.
 
As seen for the data-set covering the
entire period, the method classifies the currencies in the pre-crisis era
broadly into two
groups with one group comprising the relatively stable
currencies of developed and emerging economies.
However, during and following the crisis, the broad division separates
only a small group of frontier economy currencies (with the exception
of VEB in Periods III and IV, and ISK in Period IV) from the rest.
While the developed economies largely remain members of the same 
group even after the crisis, the relative position of most currencies
show large changes. Even among the developed economy currencies, we
observe that ISK moves away from its peers following the crisis.
This is possibly related
to the Icelandic crisis of 2008-2010 that saw a complete
collapse of its financial system~\cite{Carey2009,Sigurjonsson2011}. The 
severity of the global crisis
of 2008 is also reflected in the movement of GBP away from the neighborhood of
currencies belonging to the other developed economies in Period III,
although it subsequently returned in Period IV to the group in which
it belonged.

\section {Discussion and Conclusion} 
In this work, we have used the Jensen-Shannon divergence, that measures the                     
difference between two distributions, in 
a novel method for clustering currencies into 
similarity groups based on the statistical behavior of their
fluctuations.
In principle, one can use other 
definitions
for the distance between probability distributions, such as the total
variation distance and the Bhattacharyya
distance~\cite{Bhattacharyya1943}.
In the case of the international currency market this approach 
is particularly apt, compared to correlation-based methods, for reconstructing the hierarchical network
relating the movements
of different currencies. This is because trading in different currencies
is distributed across the globe, with activity in different locations
(focused around specific sets of currencies) peaking at different times
because of the corresponding time-zones. This is quite distinct from the
situation for a stock market
located in a specific geographical location, where trading in all
assets begins and ends at the same time, which allows one to compute the
synchronous correlations between the movements of the different assets.
In contrast, when information about trading in different currencies is pooled
together over a given 24 hour period, one may be comparing the movement of 
a particular currency at the beginning of the trading day in the location
where trading in it is most intense with that of another currency
at the close of the trading day in a different location which focuses on 
trading in the latter. This makes it essentially impossible to
obtain a truly synchronized correlation matrix for the currency market, unlike
the situation in the stock market. Thus, the network of relations between 
currencies inferred from a correlation-based approach is likely to
have inaccuracies. The technique based on measuring the similarity between
fluctuation distributions of currencies that is described here does 
not have the limitation of requiring the information for the different
assets to be acquired in a synchronized manner. It is therefore 
more likely to provide a correct description of the underlying 
hierarchical structure of the market.

Comparing the hierarchical clustering organization in the currency market 
across different periods can provide us with an indication of the
intensity of specific disruptive events in the economy of different nations.
For instance, we observe that ISK and HKD have changed their
position relative to other currencies in the dendrograms representing
hierarchical clustering of the currencies at different eras
(Fig.~\ref{fig7}). ISK lies close to the cluster of developed economy
currencies in the first two periods considered, but neighbors emerging
and frontier economy currencies in the later periods. This helps us to 
connect the atypical characteristics shown by the currency with the
effects of the major financial crisis that affected Iceland in the
latter
era. Triggered by the default of all three major privately-owned
commercial banks in Iceland in 2008, the crisis resulted in the first
systemic collapse in any advanced economy~\cite{Danielsson2009}. 
A sharp drop in the value of ISK
followed, with exchange transactions halted for weeks and the
value of stocks in the financial market collapsing. 
The crisis led to a severe economic depression in Iceland 
lasting from 2008-2010.
By contrast, HKD appears close to other developed economy currencies
in Period I, but in the neighborhood of emerging and frontier
currencies in subsequent periods. This again helps us to link the 
unusual behavior of HKD with the crisis triggered by the SARS 
epidemic of 2003 affecting mainland
China, Taiwan and large parts of Southeast Asia, that
caused extensive economic damage to Hong Kong with
unemployment hitting a record high~\cite{Chan2010}.
For the Hong Kong currency and banking system that had survived the Asian
Financial Crisis of 1997-98~\cite{Jao2001}, 
the epidemic was an unexpected shock~\cite{Siu2004}, with a net capital
outflow observed during the persistent phase of the
disease~\cite{Lee2004}. In addition, the
dominance of the service sector in the Hong Kong economy meant that
the reduction in contact following the epidemic outbreak had a large 
negative impact on the GDP~\cite{Lee2004}.
Thus, the deviation in the behavior of specific currencies from that
expected because of the macro-economic characteristic can be
traced to particular disruptive events that specifically affected
them.

To conclude, we have quantitatively measured the degree of similarity 
between different currencies by applying a distance metric on
the distribution of their fluctuations.
This has allowed us to uncover the hierarchical structure relating
movements of different currencies in the international FOREX market,
in particular, showing that currencies belonging to economies of
similar nature tend to cluster together during normal periods
in the market. More importantly, economic disruptions in a 
particular country tend to displace it from its usual position in the 
hierarchy. The severe impact of the global financial crisis of 2008 is
reflected in the large-scale disruption of the arrangement of the clusters 
that were in existence in pre-crisis times.
Thus, considering the temporal
dimension in our analysis allows us to relate particularly strong
economic shocks to changes in the relative positions of currencies in
their hierarchical clustering. We suggest that
using statistical distance between distributions characterizing the
dynamical states of the components of a complex system, to reconstruct
the hierarchical arrangement of the network of inter-relations between them, can complement
existing correlation-based methods. Indeed, in situations where information
about the different components cannot be 
obtained simultaneously
(thereby precluding the construction of a synchronous cross-correlation 
matrix), the technique proposed here may provide the means for
accurately constructing the network of interactions between the 
components.

\begin{center}
{\bf Acknowledgments}\\ 
\end{center}
We thank Anindya S. Chakrabarti, Tanmay Mitra and V. Sasidevan for helpful
suggestions. We gratefully acknowledge the assistance of Uday Kovur in the
preliminary stages of this work. This work was supported in part by the IMSc
Econophysics (XII Plan) Project and by the IMSc Centre of Excellence in Complex Systems \& Data Science, both funded by the Department of Atomic Energy,
Government of India.

\end{document}